\def\thisdate{September 4, 2015}
\documentclass[aps,prc,superscriptaddress,twoside,twocolumn,nofootinbib,showpacs]{revtex4-1}
\usepackage{amssymb}
\usepackage{amsmath}
\usepackage{graphicx}
\graphicspath{{figs/}}
\usepackage{ulem}
\usepackage[usenames, dvipsnames]{xcolor}

\newcommand*{\Mint}{M_{\text{int}}}
\newcommand*{\IC}{{\text{int}}}
\newcommand*{\mP}{\mathcal{P}}
\newcommand*{\mM}{\mathcal{M}}
\newcommand*{\mF}{\mathcal{F}}
\newcommand*{\mC}{\mathcal{C}}

\newcommand*{\scale}{{\text{sc}}}
\newcommand*{\hF}{\boldsymbol{G}}
\newcommand*{\iso}{\boldsymbol{\tau}}
\newcommand*{\RF}{F_{\textsc{r},t}}
\newcommand*{\RFF}{\mathcal{F}_{\textsc{r},t}}
\allowdisplaybreaks

\begin{document}

\title{\boldmath Preserving Local Gauge Invariance with $t$-Channel Regge Exchange}
\author{Helmut Haberzettl}
\email{helmut.haberzettl@gwu.edu} \affiliation{Institute for Nuclear Studies
and Department of Physics, The George Washington University, Washington, DC
20052, USA}
\author{Xiao-Yun Wang}
\email{xywang@impcas.ac.cn} \affiliation{Institute of Modern Physics, Chinese
Academy of Sciences, Lanzhou 730000, China} \affiliation{University of Chinese
Academy of Sciences, Beijing 100049, China}
 \affiliation{Research Center for Hadron and CSR Physics,
Institute of Modern Physics of CAS and Lanzhou University, Lanzhou 730000,
China}
\author{Jun He}
\email{junhe@impcas.ac.cn} \affiliation{Institute of Modern Physics, Chinese
Academy of Sciences, Lanzhou 730000, China}
 \affiliation{Research Center for Hadron and CSR Physics,
Institute of Modern Physics of CAS and Lanzhou University, Lanzhou 730000,
China}
\affiliation{State Key Laboratory of Theoretical Physics, Institute of
Theoretical Physics, Chinese Academy of Sciences, Beijing 100190, China}

%%%%%%%%%%%%%%%%%%%%
%%%%%%%%%%%%%%%%%%%%
\begin{abstract}
Considering single-meson photo- and electroproduction off a baryon, it is
shown how to restore \textit{local} gauge invariance that was broken by
replacing standard Feynman-type meson exchange in the $t$-channel by exchange
of a Regge trajectory. This is achieved by constructing a contact current
whose four-divergence cancels the gauge-invariance-violating contributions
resulting from all states above the base state on the Regge trajectory. To
illustrate the procedure, modifications necessary for the process $\gamma +p
\to K^+ +\Sigma^{*0}$ are discussed in some detail. We also provide the
general expression for the contact current for an arbitrary reaction.
\end{abstract}
%%%%%%%%%%%%%%%%%%%%
%%%%%%%%%%%%%%%%%%%%

%\hfill\thisdate
\date{\thisdate}
\maketitle

%%%%%%%%%%%%%%%%%%%%
\section{Introduction}

Photo- and electroproduction of mesons off baryons provide arguably the most
direct routes to information about hadronic structure. At high energies, where
multi-meson production abounds, such processes can be described economically in
terms of pomeron and Regge-trajectory exchanges~\cite{Collins,DDLN,Gribov}. At
lower energies, single-meson production provides a direct avenue for baryon
spectroscopy~\cite{KR2010}, with theoretical descriptions that attempt to model
the contributing mechanisms as detailed as possible in terms of Feynman-type
exchange processes.

\begin{figure}[b!]\centering
  \includegraphics[width=\columnwidth,clip=]{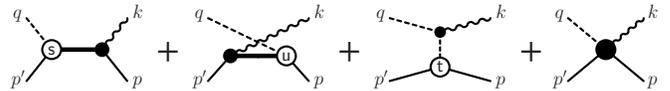}
  \caption{\label{fig:Msutc}%
  Generic diagrams with external four-momenta of the photoproduction process of
  Eq.~(\ref{eq:process}) satisfying $q+p'=k+p$. Labels $s$, $u$, and $t$ at the
  hadronic $b\to m+b'$ vertices refer to Mandelstam variables of the respective
  exchanged intermediate particles. Summations over all intermediate states
  compatible with initial and final states are implied. The right-most diagram
  depicts the contact-type interaction current. Time runs from right to left.}
\end{figure}

The present work is concerned with an intermediate-energy transition region,
where one starts within the Feynman-type picture and replaces some exchanges by
Regge trajectories in an attempt to bring the economic features of the
high-energy Regge approach to bear in the more traditional Feynman framework.
Specifically, we will apply such a hybrid framework to the generic
electromagnetic production process depicted in Fig.~\ref{fig:Msutc} of a meson
($m$) off an initial baryon ($b$) going over into a final baryon ($b'$), i.e.,
\begin{equation}
  \gamma(k)+b(p)\to m(q)+b'(p')~,
  \label{eq:process}
\end{equation}
where arguments denote the corresponding four-momenta.

%
%=======================================================
%=======================================================
%

For this single-meson production process, it is argued that replacing the
$t$-channel single-meson exchange (third diagram in Fig.~\ref{fig:Msutc}) by
the exchange of an entire Regge trajectory, would lead to a better, simpler
description of the dynamics of the process in question, in
particular, if it is dominated by small-momentum transfers~%
\cite{GLV97,VGL98,Corthals06,Corthals07,NK2010,He14,HeNP14,NY11,HC12,NG13,Wang14,Chiang,Titov,Hyodo,Toki}.
However, the good success of such hybrid approaches notwithstanding, it is well
known that this replacement destroys gauge invariance even if the underlying
Feynman formulation was gauge invariant to start with.

One widely used recipe for restoring gauge invariance is the method of
Ref.~\cite{GLV97} which basically uses the residual function of the base state
of the $t$-channel Regge trajectory as a common suppression function for all
terms of the tree-level current. Current conservation --- i.e., $k_\mu
M^\mu=0$, where $M^\mu$ denotes the current --- is achieved in this method
because one starts from a conserved tree-level current, and multiplication by a
common suppression function does not destroy this property. Even though the
method is quite successful in providing good descriptions of data in many
applications (see, for example,
Refs.~\cite{GLV97,VGL98,Corthals06,Corthals07,NK2010,He14,HeNP14,NY11,HC12,NG13,Wang14,Chiang}),
there is no dynamical foundation for it.

We point out in this context that the current-conservation condition, $k_\mu
M^\mu=0$, only implies \textit{global} gauge invariance which is little more
than charge conservation. \textit{Local} gauge invariance, i.e., the
requirement that the physical observables are invariant under local U(1)
transformations of the fields, on the other hand, implies the very existence of
the electromagnetic field~\cite{PStext}. Since global gauge invariance follows
from local gauge invariance, but not the other way around, imposing current
conservation by itself to find ways of repairing a current that was damaged by
approximations, therefore, does not imply that the damage done to the
underlying electromagnetic field is repaired as well.

We will show here how \textit{local} gauge invariance can be restored when the
$t$-channel single-meson exchange is replaced by the exchange of an entire
mesonic Regge trajectory. The method as such is not restricted to the
$t$-channel and could also be applied to a $u$-channel description in terms of
baryon Regge trajectories in a straightforward
manner.\footnote{\label{foot:schannel}%
  Technically, it could also be used for $s$-channel Reggeization, but since
  the $s$-channel contribution for a given experiment is a constant, without
  any angular dependence, it seems doubtful that there would be much point in
  doing so, even if one ignores duality issues between $s$- and $t$-channel
  processes~\cite{Collins,DDLN,Gribov}.}
The proposed mechanism is based on the necessary and sufficient conditions for
local gauge invariance formulated as generalized Ward-Takahashi identities for
the production current~\cite{Kazes1959,hh97}. These are \textit{off-shell}
conditions that automatically reduce to the familiar current-conservation
relation, $k_\mu M^\mu=0$, when taken on shell. The implementation of these
conditions results in contact-type interaction currents~\cite{hh98,hh06,hh11}
as minimal additions to a given current to restore local gauge invariance. The
method is well established within the usual Feynman picture and it has been
applied successfully to a variety of
photoprocesses~\cite{NH04,NH06,NOH06,HL08,yo08,NH09,NOH12,Huang12,HN12,HHN13,Roenchen13}.
The extension given here to include exchanges of Regge trajectories is
straightforward.

The paper is organized as follows. In the subsequent Sec.~\ref{sec:basics}, we
will recapitulate basic details of meson photoproduction within the general
field-theory approach of Haberzettl~\cite{hh97} and discuss, in particular, how
the set of generalized Ward-Takahashi identities ensures the local gauge
invariance of the production current. The Regge treatment of $t$-channel meson
exchanges considered in Sec.~\ref{sec:regge} is then immediately seen to
violate these conditions thus leading to a current that is not conserved. The
reason for this violation can be traced to the fact that higher-lying mass
states above the base state of the Regge trajectory have the wrong coupling to
the electromagnetic field. Using the residual function from the base state of
the Regge trajectory, we show then how to construct a contact current that
restores validity of the full set of generalized Ward-Takahashi identities and
therefore ensures local gauge invariance. As an illustration of the relevant
details, we treat in Sec.~\ref{sec:example} the example of the
strangeness-production reaction $\gamma+p\to K^+ + \Sigma^{*0}$. In
Sec.~\ref{sec:summary}, we will provide a summarizing discussion of the present
approach. Finally, in the Appendix, we write out the generic expressions
applicable to any single-meson production process that allow one to construct
the minimal contact currents necessary to maintain local gauge invariance.

%%%%%%%%%%%%%%%%%%%%%%%%%%%%%%%%%%%%%%%%%%%
%%%%%%%%%%%%%%%%%%%%%%%%%%%%%%%%%%%%%%%%%%%
%%%%%%%%%%%%%%%%%%%%%%%%%%%%%%%%%%%%%%%%%%%
\section{Photoproduction Basics}\label{sec:basics}

The following  description is based on the field-theoretical approach of
Haberzettl~\cite{hh97} originally developed for pion photoproduction off the
nucleon. This formalism, however, is quite generic and can be readily applied
to meson-production processes off any baryon.

The basic topological structure of the single-pion production current $M^\mu$
was given a long time ago~\cite{GellMann54} as arising from how the photon can
couple to the underlying hadronic $\pi NN$ vertex. The resulting structure
depicted in Fig.~\ref{fig:Msutc} is generic and applies to all photo- and
electroproduction processes of a single meson off a baryon. The full current
$M^\mu$, therefore, can be written generically as
\begin{equation}
  M^\mu = M^\mu_s+M^\mu_u+M^\mu_t +M^\mu_\IC~,
  \label{eq:Msuti}
\end{equation}
as indicated in Fig.~\ref{fig:Msutc}, where the indices $s$, $u$, and $t$ here
refer to the  Mandelstam variables of the respective exchanged intermediate
off-shell particle. This structure is based on topology alone and therefore
independent of the details of the individual current contributions. The first
three (polar) contributions are relatively simple; the real complication of the
problem lies in how complex the reaction mechanisms are that are taken into
account in the interaction current $\Mint^\mu$ because in principle $\Mint^\mu$
subsumes all mechanisms that do not have $s$-, $u$-, or $t$-channel poles, and
this comprises \textit{all} final-state interactions and therefore necessarily
all effects that arise from the coupling of various reaction
channels~\cite{hh97,hh11,Roenchen13}.

Here, we will ignore all of these reaction-dynamical complications and treat
the interaction current $M_\IC^\mu$ simply as a `black box' that must satisfy
certain four-divergence constraints~\cite{hh97}. If needed, one may add the
manifestly transverse contributions of the more complete
treatment~\cite{hh06,hh11} to the minimal explicit structure discussed here.

We emphasize that the particles explicitly entering all expressions here must
be physical particles. In other words, the Regge-specific implementation of the
formalism does not apply to bare particles. The corresponding propagators here
must describe physical particles, with poles at the respective physical masses,
but their structure is not limited otherwise, i.e., they may contain explicit
dressing functions or they can be simple Feynman-type propagators, however,
with physical masses, with the dressing mechanisms that gave them their
physical masses hidden in form factors. In other words, the diagrams of
Fig.~\ref{fig:Msutc} must be taken as representing the solution of the
meson-production problem and not as the Born-type bare input for a
Bethe-Salpeter- or Dyson-Schwinger-type reaction equation.

Also, for the purpose of gauge invariance, the only relevant intermediate
states in the $s$-, $u$-,  and $t$-channel diagrams of Fig.~\ref{fig:Msutc} are
those where the photon does not initiate a transition  (since transition
currents are transverse), i.e., where the states before and after the photon
interacts are the same particle with non-zero charge. Thus, for the present
purpose, without lack of generality, we may ignore all diagrams and
intermediate states that do not contribute to the four-divergence of the
production current $M^\mu$.

\begin{figure}[t!]\centering
  \includegraphics[width=.35\columnwidth,clip=]{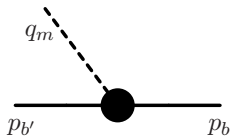}
  \caption{\label{fig:MBBvertex}%
  Generic vertex $F(p_{b'},p_b)$ for  $b\to m+b'$ with associated momenta. The
  meson momentum $q_m=p_b-p_{b'}$ is given by four-momentum conservation across
  the vertex.}
\end{figure}

As a consequence, with this restriction,  all three hadronic vertices in
Fig.~\ref{fig:Msutc} describe the same three-point vertex $b\to m+b'$, for
which we will use the notation $F(p_{b'},p_b)$, where the arguments here are
the incoming and outgoing baryon momenta, as depicted in
Fig.~\ref{fig:MBBvertex}. The vertex notation $F$ subsumes all coupling
operators and isospin dependence, etc., and depending on the specific reaction,
it may also carry Lorentz indices [see Eq.~(\ref{eq:modelvertex}) below, and
also the example in Sec.~\ref{sec:example}.] The three kinematic situations in
which this vertex appears in Fig.~\ref{fig:Msutc} are then uniquely identified
by the Mandelstam variables of the exchanged intermediate hadron,
\begin{subequations}\label{eq:Mandelstam}
\begin{align}
s&=(p+k)^2=(p'+q)^2~,
\\
u&=(p'-k)^2=(p-q)^2~,
\\
t&=(q-k)^2=(p-p')^2~,
\end{align}
\end{subequations}
and we will use
\begin{equation}
  F_t=F(p',p)\,,~~F_u=F(p'-k,p)\,,~~F_s= F(p',p+k)
\end{equation}
to abbreviate the corresponding vertices, and generically write $F_x$ for
$x=s,u,t$.

%
%=======================================================
%=======================================================
%

\subsection{Generalized Ward-Takahashi identities}

First, to set the stage for the Regge treatment, we will recapitulate how the
local gauge-invariance requirements differ from mere current conservation,
i.e., global gauge invariance.

To preserve \textit{local} gauge invariance for the photoprocess
(\ref{eq:process}) the following set of \textit{off-shell} four-divergence
relations need to be satisfied~\cite{hh97,hh06,hh11}. At the very base are the
Ward-Takahashi identities (WTI)~\cite{Ward,Takahashi} for the individual
electromagnetic currents $J^\mu$ of mesons (index $m$) and baryons (indices
$b'$ or $b$),
\begin{subequations}\label{eq:WTI}
\begin{align}
  k_\mu J_m^\mu(q,q-k) &= \Delta_{m}^{-1}(q) Q_{m}-  Q_{m} \Delta_{m}^{-1}(q-k)~,
  \\
  k_\mu J_{b'}^\mu(p',p'-k) &= S_{b}^{-1}(p') Q_{b'}-  Q_{b'} S_{b}^{-1}(p'-k)~,
  \\
  k_\mu J_b^\mu(p+k,p) &= S_{b}^{-1}(p+k) Q_{b}-  Q_{b} S_{b}^{-1}(p)~,
\end{align}
\end{subequations}
where $\Delta_m(q)$, $S_{b'}(p')$ and $S_{b}(p)$ are the respective propagators
for the meson and baryons, with arguments providing their four-momenta, and
$Q_m$, $Q_{b'}$, and $Q_b$ denoting their associated charge operators. The
photoproducton current $M^\mu$ of Eq.~(\ref{fig:Msutc}) must satisfy the
generalized WTI (gWTI)~\cite{Kazes1959,hh97},
\begin{align}
k_\mu M^\mu &= \Delta_{m}^{-1}(q) Q_{m} \Delta_{m}(q-k)\, F_t
\nonumber\\
&\mbox{}\qquad
+ S_{b'}^{-1}(p') Q_{b'} S_{b'}(p'-k)\, F_u
\nonumber\\
&\mbox{}\qquad\quad - F_s\, S_b(p+k) Q_b S_b^{-1}(p)~,
\label{eq:gWTI}
\end{align}
and, finally, the interaction current $\Mint^\mu$ needs to satisfy the
condition
\begin{align}
  k_\mu \Mint^\mu &= Q_{m}\, F_t + Q_{b'}\, F_u
  - F_s\,Q_b~.
  \label{eq:gWTIint}
\end{align}
In view of the isospin dependence of the vertices, charge operators and
vertices do not commute. Note that the right-hand side vanishes here
identically if all vertices are replaced by simple coupling constants for we
have then $F_x \to g \iso$, where $g$ is the coupling constant and $\iso$
generically denotes the isospin operator of the vertex, and hence $Q_m\iso
+Q_{b'}\iso - \iso Q_b\equiv 0$ provides charge conservation across the
photoprocess~\cite{hh97}. In a manner of speaking, therefore,
Eq.~(\ref{eq:gWTIint}) amounts to the formulation of the effective charge
difference across the reaction in the presence of hadronic vertices with
structure.

It is of paramount importance here that all three four-divergence equations are
off-shell relations, and that the off-shellness is a necessary requirement for
local gauge invariance~\cite{hh97} since it ensures that the (off-shell)
current $M^\mu$ provides the correct, consistent contributions to gauge
invariance even if it is embedded as an off-shell subprocess in a larger
process (for example, electromagnetic production of two or more
mesons~\cite{NOH06,FB20}).

With the off-shell WTIs (\ref{eq:WTI}) and (\ref{eq:gWTI}) given, global gauge
invariance follows trivially by taking the respective on-shell matrix elements,
with the inverse propagators in the four-divergences (\ref{eq:WTI}) and
(\ref{eq:gWTI}) then ensuring that the four-divergences vanish; in particular,
\begin{equation}
  k_\mu M^\mu =0\qquad \text{(on shell)}~.
\label{eq:CurrConserved}
\end{equation}
To be sure, this is a necessary condition the physical production current needs
to satisfy that follows trivially from local gauge invariance, however, this
on-shell restriction by itself contains no information that allows one to
meaningfully `guess' at a nontrivial structure for $M^\mu$. Thus, it should not
be used as a starting point for restoring gauge invariance destroyed by
approximations.

The proper starting point should be the set of off-shell equations
(\ref{eq:WTI}), (\ref{eq:gWTI}), and (\ref{eq:gWTIint}). One easily sees here
that only two --- any two --- of these conditions are necessary to ensure the
validity of the respective third equation. % (see also Sec.~\ref{sec:tCh} below).
For the practical purpose of restoring gauge invariance, it is easiest to work
with Eqs.~(\ref{eq:WTI}) and (\ref{eq:gWTIint}). In any microscopic formulation
of photoprocesses, the single-hadron WTIs (\ref{eq:WTI}) are a given from the
start. Therefore, to obtain the gWTI (\ref{eq:gWTI}) for the full production
current $M^\mu$ and thus ensure the preservation of local gauge invariance, one
needs to construct an interaction current $\Mint^\mu$ that satisfies
Eq.~(\ref{eq:gWTIint}). Note, in particular, that the structure of this
equation does not change even if the external hadrons are on shell, and thus
--- quite in contrast to the current-conservation condition (\ref{eq:CurrConserved})
--- \textit{even its on-shell limit provides a nontrivial constraint} that ensures
that the on-shell result (\ref{eq:CurrConserved}) is a consequence of
\textit{local} gauge invariance and not just mere global gauge invariance.

\subsubsection{$t$-Channel contribution}\label{sec:tCh}

To see what needs to be done to restore gauge invariance in the Regge case, let
us first look at how the usual $t$-channel term as depicted by the third
diagram in Fig.~\ref{fig:Msutc} contributes to upholding local gauge
invariance.

Using the momenta of the diagram and stripped of all unnecessary factors, it
reads
\begin{equation}
  M_t^\mu = J^\mu_m (q,q-k) \Delta_m(q-k) F_t~,
  \label{eq:Feynman}
\end{equation}
and its four-divergence is given by
\begin{equation}
  k_\mu M_t^\mu =  \Delta_{m}^{-1}(q) Q_{m} \Delta_m(q-k) F_t
  -  Q_{m} F_t~.
  \label{eq:kmuMt}
\end{equation}
The first term on the right hand side is precisely the first term appearing on
the right-hand side of the gWTI~(\ref{eq:gWTI}); the second term involving only
the vertex, but no propagator, is canceled by the first term on the right-hand
side of the interaction-current condition~(\ref{eq:gWTIint}). Similar
cancellations happen for the respective contributions from all three polar
current contributions and this cancellation mechanism ensures the validity of
the full gWTI --- and thus of local gauge invariance
--- once Eqs.~(\ref{eq:WTI}) and (\ref{eq:gWTIint}) are satisfied.

It is this cancellation mechanism that will be exploited in the subsequent
Regge treatment.

%%%%%%%%%%%%%%%%%%%%%%%%%%%%%%%%%%%%%%%%%%%
%%%%%%%%%%%%%%%%%%%%%%%%%%%%%%%%%%%%%%%%%%%
%%%%%%%%%%%%%%%%%%%%%%%%%%%%%%%%%%%%%%%%%%%
\section{Gauge-invariant Regge treatment}\label{sec:regge}

First, let us write the hadronic $b\to m+b'$ vertex (see
Fig.~\ref{fig:MBBvertex}) as\footnote{For a more general description of the
vertex, see discussion in Appendix \ref{app:beyond}.}
\begin{equation}
  F(p_{b'},p_b) = \hF(q_m)\iso\, f(q_m^2,p_{b'}^2,p_b^2)~,
  \label{eq:modelvertex}
\end{equation}
where the outgoing meson four-momentum $q_m$ is given by $q_m = p_b-p_{b'}$ in
terms of the incoming and outgoing baryon momenta. The operator $\hF$
describing the coupling structure of the vertex subsumes all strength
parameters, masses, signs, etc.; in the simplest case it is just a constant,
but in more complicated cases it contains derivatives of the outgoing meson
field which lead to the $q_m$ dependence. The extended structure of the vertex
is given by the scalar form factor $f$ normalized as
\begin{equation}
  f(M_m^2,M_{b'}^2,M_b^2) =1~,
  \label{eq:norm}
\end{equation}
where the squared momenta of (\ref{eq:modelvertex}) sit on there respective
mass shells. The operator $\iso$ summarily describes the isospin dependence of
the vertex, with relevant indices suppressed. Combined with the respective
charge operators $Q$ for the three legs of the vertex, one obtains~\cite{hh97}
\begin{equation}
  Q_m\iso =e_m~,\quad Q_{b'}\iso=e_{b'}~,\quad \iso Q_b=e_b~,
\end{equation}
where
\begin{equation}
  e_m + e_{b'} -e_b =0
  \label{eq:conscharge}
\end{equation}
provides charge conservation across the reaction. Taken in an appropriate
isospin basis, the charge-isospin operators $e_m$, $e_{b'}$, and $e_b$ are
equal to the respective charges of the individual legs.

We will need only on-shell kinematics here where all external hadron legs of
Fig.~\ref{fig:Msutc} sit on their respective mass shells. The form factors
associated with the vertices $F_x$, $x=s,u,t$, for these cases are
\begin{subequations}
\begin{align}
 f_s(s) &= f(M_m^2,M_{b'}^2,s)~,
 \\
 f_u(u) &= f(M_m^2,u,M_b^2)~,
 \\
 f_t(t) &= f(t,M_{b'}^2,M_b^2)~,
\end{align}
\end{subequations}
where the Mandelstam variables (\ref{eq:Mandelstam}) are used. The $t$-channel
vertex, in particular, then reads
\begin{equation}
  F_t= F(p',p) =  \hF(q-k)\,\iso\,f_t(t)~,
\end{equation}
and for the corresponding meson-exchange propagator, we may write without lack
of generality
\begin{equation}
  \Delta_m(q-k) = \frac{N_m(q-k)}{t-M_m^2}~,
  \label{eq:tpole}
\end{equation}
where the pole at $t=M_m^2$ was pulled out explicitly and the residual
numerator $N_m(q-k)$ defined by this relation may describe dressing effects
and/or the coupling structure of the propagator. In the simplest cases, $N_m$
equals unity for pseudoscalar mesons and $-g^{\beta\alpha}$ for vector mesons
(in Feynman gauge), for example.

Standard Reggeization of the $t$-channel meson exchange  corresponds to the
replacement~\cite{GLV97,VGL98,Chiang,Titov,Hyodo,Toki,Corthals06, Corthals07}
\begin{equation}
 \frac{1}{t-M_m^2} f_t(t) \to \mP_m(t)~,
 \label{eq:replaceRegge}
\end{equation}
where $\mP_m(t)$ is the Regge-trajectory propagator appropriate for this
particular meson exchange. By construction (see details
Sec.~\ref{sec:resfunction}), it contains poles at higher-lying meson masses
along this particular trajectory, in addition to the primary pole at the base
of the trajectory at $t=M_m^2$ of $(\ref{eq:tpole})$. Moreover, the residue at
this primary pole,
\begin{equation}
  \lim_{t\to M_m^2} (t-M_m^2) \mP_m(t) =1~,
  \label{eq:limP}
\end{equation}
is exactly the same as that of the left-hand side of (\ref{eq:replaceRegge}).
The residual function,\footnote{%
     It is this residual function that was used in Ref.~\cite{GLV97} as an overall
     multiplicative factor for their gauge-invariance-restoring recipe. }
\begin{equation}
\mF_t(t)=(t-M_m^2)\,\mP_m(t)~,
\label{eq:residualfct}
\end{equation}
thus, is finite and normalized to unity at $t=M_m^2$, just like the usual
$t$-channel form factor $f_t(t)$. Details of $\mF_t$ will be given in the
subsequent Sec.~\ref{sec:resfunction}.

The Reggeized $t$-channel current now reads
\begin{equation}
M_t^\mu \to\mM_t^\mu = J^\mu_m (q,q-k)\,\Delta_m(q-k)\,\RF(p',p)~,
    \label{eq:tReggeMod}
\end{equation}
where
\begin{equation}
\RF(p',p)= \hF(q-k)\,\iso\,\mF_t(t)
\end{equation}
describes the Reggeized vertex, with the corresponding four-divergence given by
\begin{align}
  k_\mu \mM^\mu_t &=\Delta_m^{-1}(q)Q_m\,\,\Delta_m(q-k)\,\RF
    -Q_m\, \RF~.
    \label{eq:gaugeRviol}
\end{align}
The first term on the the right-hand side with the inverse meson propagator
depending on the external (outgoing) meson momentum vanishes on-shell and thus
provides an acceptable contribution for gWTI in analogy to Eq.~(\ref{eq:gWTI}).
The second term, however, has no counterpart in the four-divergence
(\ref{eq:gWTIint}) and thus violates local gauge invariance (and therefore
obviously also global gauge invariance).

This violation comes about because in the Regge treatment \textit{all}
particles on the trajectory are taken to couple to the photon with the
\textit{same} current $J^\mu_m$ as the primary base state, whereas if one were
to incorporate these contributions via Feynman-type exchange mechanisms, each
of the higher-lying states would couple transversely to the photon because the
corresponding currents are transition currents for the transition from
intermediate higher-mass states to the lower-mass primary base state, which is
the final meson state of the reaction, and such transverse transition currents
would not contribute to the four-divergence.

Clearly, to restore local gauge invariance, Regge treatment of the $t$-channel
by itself is not enough --- one \textit{must} also Reggeize the interaction
current $\Mint^\mu$ so that its four-divergence will provide the necessary
cancellation of the offending contribution in (\ref{eq:gaugeRviol}), thus in
essence restoring the transversality of these contributions with higher mass.
In other words, to preserve local gauge invariance, one must apply the
Reggeization process consistently across all elements of the production current
$M^\mu$. Since $t$-channel-type exchanges also contribute (as off-shell
processes) to the internal mechanisms of $\Mint^\mu$, an appropriate
Reggeization of such internal exchanges should provide the cancellation for the
offending term in Eq.~(\ref{eq:gaugeRviol}).

Obviously then, treating Regge consistently with local gauge invariance simply
entails consistently replacing the usual $t$-channel vertex $F_t$ by the
Reggeized vertex $\RF$ everywhere. In addition to the Reggeized $t$-channel
current (\ref{eq:tReggeMod}), this also requires modification of the contact
current,
\begin{equation}
F_t\to \RF\text{:}\qquad  \Mint^\mu \to \mM_\IC^\mu~,
\end{equation}
such that the Reggeized contribution from the corresponding four-divergence,
\begin{equation}
  k_\mu \mM_\IC^\mu = Q_m\, \RF + Q_{b'}\, F_u
- F_s\,Q_b~,
  \label{eq:kMintRegge}
\end{equation}
now cancels the previously gauge-invariance-violating term from
(\ref{eq:gaugeRviol}).

The resulting Reggeized photoamplitude,
\begin{equation}
  M^\mu \to \mM^\mu = M_s^\mu+M_u^\mu + \mM^\mu_t +\mM^\mu_\IC~,
\end{equation}
then, by construction, satisfies the appropriate gWTI,
\begin{align}
k_\mu \mM^\mu &= \Delta_m^{-1}(q) Q_m \Delta_m(q-k)\, \RF
\nonumber\\
&\mbox{}\qquad + S_{b'}^{-1}(p') Q_{b'} S_{b'}(p'-k)\, F_u
\nonumber\\
&\mbox{}\qquad\quad - F_s\, S_b(p+k) Q_b S_b^{-1}(p)~,
\end{align}
and thus is fully consistent with local gauge invariance.

The construction of the Reggeized contact current $\mM^\mu_\IC$ that produces
the correct four-divergence (\ref{eq:kMintRegge}) from the Reggeized vertex
$\RF$ follows exactly along the same lines as those given for un-Reggeized
contact currents $\Mint^\mu$~\cite{hh06}. The procedure is straightforward, and
we provide the corresponding generic expressions for the minimal interaction
current that restores local gauge invariance in the Appendix. However, to
understand how it works, it might be more illuminating to consider an example.
To this end, we discuss in Sec.~\ref{sec:example} a strangeness-production
process with a Kroll-Ruderman-type~\cite{Kroll1954} bare contact current.

\subsection{Regge residual function}\label{sec:resfunction}

To provide explicit expressions for the residual function
(\ref{eq:residualfct}), it is convenient to rewrite  the standard expressions
for positive- and negative-signature Regge propagators given in
Refs.~\cite{Collins,GLV97} to obtain the unified form
\begin{align}
  \mF_t(t)
  = \left(\frac{s}{s_\scale}\right)^{\alpha_i(t)}
  \frac{N\big[\alpha_i(t);\eta\big]}{\Gamma\big(1+\alpha_i(t)\big)}
  \frac{\pi\alpha_i(t)}{\sin\big(\pi\alpha_i(t)\big)}~,
  \label{eq:FRegge}
\end{align}
where the functions
\begin{equation}
  \alpha_i(t) = \alpha_i'\,(t-M_i^2)~, \qquad \text{for}~~i=0,1~,
\end{equation}
are related to the usual Regge trajectories by
\begin{equation}
\alpha_\zeta(t) = \begin{cases}
\alpha_0(t)~, &\text{for}~~ \zeta=+1~,
\\
1+ \alpha_1(t)~, &\text{for}~~ \zeta=-1~.
\end{cases}
\end{equation}
Here, the signature for pseudoscalar mesons is $\zeta=+1$ (corresponding to
$i=0$) and $\zeta=-1$ (corresponding to $i=1$) for vector mesons. The masses
$M_i$ here are the lowest masses at the bases of the respective trajectories,
with their slopes given by $\alpha'_i$. For these base states, at $t=M_i^2$,
the residual function thus is given by a manageable $0/0$ situation.

Even though $\mF_t$ is also $s$-dependent analytically through the scale factor
$\left(s/s_\scale\right)^{\alpha_i(t)}$, this is irrelevant for our purposes
since for a given experiment, $s$ is fixed, and we may consider $\mF_t$ as a
function of $t$ for fixed $s$. The exponential scale factor suppresses the
Regge contribution for $s>s_\scale$ for (negative) physical values of $t$; the
scale parameter $s_\scale$ usually is chosen as $s_\scale=1$\,GeV.

The signature function $N$ appears here as
\begin{equation}
 N[\alpha_i(t);\eta] = \eta + (1-\eta)e^{-i\pi\alpha_i(t)}~,
\label{eq:fitphase}
\end{equation}
where $\eta$ is a real parameter whose three standard values are
\begin{equation}
  \eta =\begin{cases}
    \frac{1}{2}~, & \text{pure-signature trajectories}~,
    \\
    0~, & \text{add  trajectories: rotating phase}~,
    \\
    1~, & \text{subtract trajectories: constant phase}~.
  \end{cases}
\end{equation}
In the pure-signature case, for $\eta=1/2$, $N$ vanishes for every odd integer
value of $\alpha_i(t)$, thus leaving only the even integer values to produce
poles in (\ref{eq:FRegge}). This corresponds to  even and odd angular momenta,
\begin{equation}
  \alpha_+ = 0,2,4,\ldots
  \quad\text{and}\quad
  \alpha_- = 1,3,5,\ldots~,
\end{equation}
associated with the states along the respective positive- or negative-signature
trajectories. Equation (\ref{eq:fitphase}) also subsumes treatment of strongly
degenerate trajectories~\cite{Collins,GLV97}, where the rotating phase
($\eta=0$) results from adding degenerate trajectories and the constant phase
($\eta=1$) arises from subtracting them. Which case applies is largely
determined semi-phenomenologically by $G$-parity arguments~\cite{GLV97}.

Going beyond these standard cases, since the signature function is largely
phenomenological anyway, one may consider $\eta$ as a convenient interpolating
fit parameter for optimizing the description of data for the value range $0 \le
\eta \le 1$. Note that $\exp(-i\pi\alpha_i)$ in (\ref{eq:fitphase}) is $+1$ at
the poles of the primary trajectory and $-1$ at the poles of the added or
subtracted secondary (degenerate) trajectory. Hence, taking into account the
minus sign arising from the negative slope of the denominator sine function in
(\ref{eq:FRegge}) at those secondary poles, this effectively changes the
coupling strength for the latter exchange by the factor $(1-2\eta)$ that can
vary between $+1$ and $-1$; it is positive or negative depending on whether its
degeneracy effect is more additive or subtractive, respectively. The coupling
strength of the primary trajectory remains unchanged. Clearly, if the
strong-degeneracy hypothesis is warranted for a particular application, fitted
values of $\eta$ should come out close to either $0$ or $1$.

At the base of the trajectories, one has
\begin{equation}
N[\alpha_i(M_i^2);\eta]=1
\end{equation}
for any value of $\eta$, thus ensuring the validity of the necessary condition
\begin{equation}
  \mF_t(M_i^2) =1~,
  \label{eq:FReggenormgen}
\end{equation}
for both $i=0,1$ for the residue of the corresponding Regge propagators. The
fact that the Regge residue function $\mF_t$ thus preserves the normalization
of the standard form factor $f_t$ is crucial for the construction of the
gauge-invariance-preserving contact current, as will be seen explicitly in the
following example.

\section{\boldmath Example: $\gamma +p \to K^+ + \Sigma^{*0}$}\label{sec:example}

In this reaction only the incoming proton and the outgoing kaon carry charge.
Hence, extracting the isospin operators from the respective vertices, the
relevant charge parameters are (in an appropriate isospin basis)
\begin{equation}
  Q_{b'}\iso\to e_\Sigma =0~,\quad Q_m\iso\to e_K=e~,\quad \iso Q_b \to e_p = e~,
\end{equation}
where $e$ is the fundamental charge unit, and charge conservation obviously
reads
\begin{equation}
e_\Sigma+ e_K = e_p
\quad\text{or}\quad
 e_K = e_p
~.
\end{equation}
Hence, as far as gauge invariance is concerned, only $s$- and $t$-channels and
a contact term contribute. It suffices to consider this as an on-shell process
if the corresponding un-Reggeized amplitude is constructed already such that it
obeys the appropriate gWTI. Moreover, we can ignore all possible resonance
contributions and other meson exchanges since they do not contribute to the
four-divergence (for a more complete discussion, see Ref.~\cite{He14}). The
only relevant exchange particles are the proton (with mass $M_N$) in the
$s$-channel and the kaon $K^+$ (with mass $M_K$) in the $t$-channel.

The $p\to K^+ \Sigma^{*0}$ vertices for the $s$- and $t$-channel terms are
given by~\cite{He14}
\begin{subequations}
\begin{align}
F_s \to F_s^\nu&=g \iso\,q^\nu f_s(s)~,
\\
F_t\to F_t^\nu &=g \iso\,(q-k)^\nu f_t(t)~,
\end{align}
\end{subequations}
with scalar form factors $f_x$ ($x=s,t$) normalized as
\begin{equation}
  f_s(M_N^2)=1
  \quad\text{and}\quad
  f_t(M_K^2)=1~.
  \label{eq:fnormalize}
\end{equation}
The constant $g$ subsumes all coupling constants, mass factors, signs, etc.,
$\iso$ generically describes the isospin dependence, and $q^\nu$ and
$(q-k)^\nu$ are the operators for $s$- and $t$-channel, respectively, providing
coupling to the spin-3/2 Rarita-Schwinger spinor of the outgoing $\Sigma^{*0}$
baryon.

The resulting current reads
\begin{equation}
  M^{\nu\mu} = M_s^{\nu\mu} + M_t^{\nu\mu} + M^{\nu\mu}_\IC~,
\end{equation}
where the Lorentz indices $\mu$ and $\nu$ connect to the incoming photon state
and the outgoing Rarita-Schwinger spinor, respectively. Assuming validity of
the single-particle WTI for the proton and the kaon (which are trivially true),
$M^{\nu\mu}$ is locally gauge invariant, according to (\ref{eq:gWTIint}), if
the interaction current satisfies
\begin{align}
  k_\mu M^{\nu\mu}_\IC
  &=Q_K F^\nu_t - F^\nu_s Q_p
  \nonumber\\
  &=e_K g\, (q-k)^\nu f_t - e_p g\, q^\nu f_s~.
  \label{eq:kMintSigma}
\end{align}
Then, explicitly writing out the $t$-channel contribution,
\begin{align}
  M_t^{\nu\mu}
  &= \frac{(2q-k)^\mu Q_K }{t-M_K^2} \, F^{\nu}_t
\nonumber\\
  &= e_K g\, \,\frac{(2q-k)^\mu }{t-M_K^2} (q-k)^\nu  f_t~,
\end{align}
we see that its (on-shell) four-divergence contribution,
\begin{equation}
  k_\mu M_t^{\nu\mu} =-Q_K F^\nu_t=  - e_K g\,  (q-k)^\nu f_t~,
  \label{eq:Mnumut}
\end{equation}
is canceled by the $t$-channel term in (\ref{eq:kMintSigma}). A similar finding
for the $s$-channel shows that the validity of (\ref{eq:kMintSigma}) is both
necessary and sufficient for making the current $M^{\nu\mu}$ locally gauge
invariant.

In the structureless limit, when all form factors are unity, the bare
interaction current $m^{\nu\mu}_c$ also must satisfy the analog of
(\ref{eq:kMintSigma}), i.e.,
\begin{align}
  k_\mu m^{\nu\mu}_c &=e_K g\, (q-k)^\nu  - e_p g\, q^\nu
  \nonumber\\
  &=k_\mu \left(-e_K g \,g^{\nu\mu}\right)~,
\end{align}
which shows that the minimal interaction current is given by
\begin{equation}
 m^{\nu\mu}_c = -e_K g \,g^{\nu\mu}~.
\end{equation}
This is precisely the contact current resulting from the usual four-point
contact Lagrangian for the present process. This result is seen here to be an
immediate consequence of local gauge invariance.

To construct the corresponding minimal interaction current, we adapt the
generic expression (\ref{eq:MintGeneric}) provided in the Appendix to the
present case and obtain
\begin{equation}
  M_\IC^{\nu\mu} = -e_Kg \, g^{\nu\mu} f_t + g\,  q^\nu C^\mu~ .
  \label{eq:Mintnumu}
\end{equation}
The auxiliary contact current,
\begin{align}
  C^\mu &= -e_K (2q-k)^\mu \frac{f_t-1}{t-M_K^2}f_s - e_p (2p+k)^\mu  \frac{f_s-1}{s-M_N^2}f_t
  \nonumber\\
  &\mbox{}\qquad
  +\hat{A}(s,t) \left(1-f_t\right)\left(1-f_s\right)
  \nonumber\\
  &\mbox{}\qquad\qquad
  \times\left[e_K\frac{(2q-k)^\mu}{t-M_K^2}+ e_p \frac{(2p+k)^\mu}{s-M_N^2}\right]~,
  \label{eq:Caux}
  \end{align}
follows from Eq.~(\ref{eq:CauxGeneric}). It was derived from imposing local
gauge-invariance requirements in the presence of vertices with
structure~\cite{hh97,hh06}. In view of the normalizations
(\ref{eq:fnormalize}), this current is manifestly nonsingular. The function
$\hat{A}(s,t)$ in front of the manifestly transverse term here is a
phenomenological (complex) function that must vanish at high energies, but
otherwise can be freely chosen to improve fits to the data.

It is now a trivial exercise to show that
\begin{equation}
  k_\mu C^\mu = e_K f_t - e_p f_s
\end{equation}
and thus the interaction current (\ref{eq:Mintnumu}) indeed provides the
correct four-divergence (\ref{eq:kMintSigma}) to ensure local gauge invariance.

We emphasize in this context that the contact-type interaction current
constructed here provides only the \textit{minimal} structure necessary for
maintaining local gauge invariance. If the physics of the problem should make
it necessary to consider additional current contributions, they can only arise
from additional manifestly transverse currents and thus do not contribute when
taking the four-divergence of the current.

\subsubsection{Regge-trajectory exchange}

To Reggeize the $K^+$ exchange of the present example, the explicit expression
in analogy to (\ref{eq:replaceRegge}) reads~\cite{GLV97}
\begin{align}
  \frac{f_t}{t-M_K^2} \to \frac{\mF_t}{t-M_K^2}~,
  \label{eq:KaonReggePole}
\end{align}
with the residual function given by (\ref{eq:FRegge}) for $i=0$, where
\begin{equation}
  \alpha_0(t) = \frac{t-M_K^2}{\Delta t_K}
\end{equation}
is the kaonic Regge trajectory, with slope
\begin{equation}
  \alpha'_0 = \frac{1}{\Delta t_K} = 0.7\,\text{GeV}^{-2}~,
\end{equation}
which puts the Regge states at
\begin{equation}
  t\to t_n=M_K^2 +n \Delta t_K~,\quad\text{for}~n=0,1,2,\ldots
\end{equation}
For pure pseudoscalar signature ($\zeta=+1~\Rightarrow~\eta=1/2$), only even
values are realized on the trajectory; for all other values of $\eta$, all
states contribute.

The Reggeized $t$-channel current reads now
\begin{equation}
  M_t^{\nu\mu} \to \mM_t^{\nu\mu}
  = e_K g \,\frac{(2q-k)^\mu}{t-M_K^2} (q-k)^\nu  \mF_t~,
\end{equation}
with the associated modified interaction current
\begin{equation}
  M_\IC^{\nu\mu} \to \mM_\IC^{\nu\mu} = -e_K g\, g^{\nu\mu} \mF_t +g\,  q^\nu \mC^\mu
\end{equation}
and modified auxiliary current
\begin{align}
  C^\mu \to \mC^\mu &= -e_K (2q-k)^\mu \frac{\mF_t-1}{t-M_K^2}f_s
  \nonumber\\
  &\mbox{}\quad - e_p (2p+k)^\mu \frac{f_s-1}{s-M_N^2}\mF_t
  \nonumber\\
  &\mbox{}\quad
  +\hat{A}(s,t)\left(1-f_t\right)\left(1-f_s\right)
  \nonumber\\
  &\mbox{}\qquad
  \times \left[e_K\frac{(2q-k)^\mu}{t-M_K^2}+ e_p \frac{(2p+k)^\mu}{s-M_N^2}\right]~.
  \label{eq:Ccontact1}
  \end{align}
Despite the Reggeization of the $t$-channel form factor, because of the limit
(\ref{eq:FReggenormgen}), this current is still nonsingular as far as the
primary propagator singularities here are concerned. Note in this respect that
there is no reason to replace $f_t$ by $\mF_t$ in the last term since this
current piece is manifestly transverse and  does not contribute to the
four-divergence. However, no harm would result if one did replace it since the
difference can be absorbed in redefining $\hat{A}$.

The auxiliary current $\mC^\mu$ now does have higher-mass singularities at
unphysical $t>0$ from the Regge trajectory but those are necessary to
compensate the corresponding higher-mass contributions from the $t$-channel
exchange which have the wrong electromagnetic coupling that led to the
violation of gauge invariance.

It is obvious now that the Reggeized production current for this process,
\begin{equation}
  M^{\nu\mu} \to \mM^{\nu\mu} = M_s^{\nu\mu} + \mM_t^{\nu\mu} + \mM^{\nu\mu}_\IC~,
\end{equation}
by construction does indeed satisfy the generalized Ward-Takahashi identity for
this process and thus provides a conserved current,
\begin{equation}
  k_\mu  \mM^{\nu\mu} = 0\qquad \text{(on shell)}~,
\end{equation}
as a matter of course.

%%%%%%%%%%%%%%%%%%%%%%%%%%%%%%%%%%%%%%%%%%%
%%%%%%%%%%%%%%%%%%%%%%%%%%%%%%%%%%%%%%%%%%%
%%%%%%%%%%%%%%%%%%%%%%%%%%%%%%%%%%%%%%%%%%%
\section{Summary and Discussion}\label{sec:summary}

We have considered here a mechanism to repair gauge invariance broken by
Reggeization of $t$-channel meson exchanges in single-meson photoproduction off
a baryon. Consistent with the underlying field-theoretical foundations of such
processes~\cite{hh97}, we have argued that this must be done by constructing
contact-type interaction currents whose four-divergence compensates for the
wrong coupling to the electromagnetic field of higher-mass contributions of the
Regge trajectory that is responsible for the violation of gauge invariance. The
construction principle was based on the underlying generalized Ward-Takahashi
identities whose validity ensure local gauge invariance.

We emphasize once more in this respect that mere (on-shell) current
conservation, $k_\mu M^\mu=0$, is not very helpful as a starting point for
repairing gauge-invariance violations. As argued, the goal of any repair
mechanism must be the construction of an interaction current $M^\mu_\IC$ that
satisfies the crucial four-divergence condition (\ref{eq:gWTIint}) for this
interaction current. The resulting local gauge-invariance property will then
automatically ensure a conserved on-shell current $M^\mu$.

The present way of maintaining local gauge invariance in terms of a Regge form
factor $\mF_t$ to replace the usual $t$-channel cutoff function $f_t$ shows
that when viewed from the Feynman perspective, the Regge approach basically can
be understood as a prescription for the functional form of the $t$-channel form
factor. Numerical test show that at (negative) physical $t$ (and fixed $s$),
the main features of $\mF_t$ that survive are the exponential scale factor and
the phase function,
\begin{equation}
  S_t(t)=\left(\frac{s}{s_\scale}\right)^{\alpha_i(t)}N\big[\alpha_i(t);\eta\big]~.
  \label{eq:ReggeScale}
\end{equation}
This exponential function falls off faster than any power-law form factor and
thus compared to a conventional phenomenological form factor drastically cuts
out the high-$|t|$ (i.e., backward-angle) scattering contributions.

The onset of the `Regge regime' is oftentimes very much under debate in
practical applications, in particular, if Regge exchanges are employed at
intermediate-energy ranges within hybrid approaches as discussed here that mix
Regge with the traditional Feynman picture. In this situation, it seems natural
to consider mechanisms for smooth transitions into that regime~\cite{Toki}. An
interpolating mechanism like $\RFF=\mF_t\,R+f_t\,(1-R)$, for example, that
determines an effective $t$-channel form factor $\RFF$ somewhere between its
non-Regge ($f_t$) and Regge ($\mF_t$) limits in terms of an ($s$- and
$t$-dependent) interpolating function $R$ can be fine-tuned to the requirements
of particular applications~\cite{Toki,NK2010}. Hence, fitting the interpolation
parameters to experimental data lets the data `decide' whether Regge exchanges
should be necessary or not for a particular process at a particular photon
energy. Since this would take much of the contention out of the debate, we
strongly advocate employing such interpolation schemes. This may be especially
advisable for energy ranges where details of baryon-resonance structure may
still play a role. Clearly, the procedure outlined here is not affected by such
an interpolation scheme since $\RFF$ is normalized to unity by construction and
may thus be used for building a contact current, just like $f_t$ or $\mF_t$.

A similarly useful interpolation procedure is provided by the $\eta$-dependence
of the signature function $N\big[\alpha_i(t);\eta\big]$ of
Eq.~(\ref{eq:fitphase}) that allows for the smooth transition from the
pure-signature case to the two limiting cases of adding or subtracting
degenerate trajectories and thus, again, lets the data decide which description
is better suited for a given application.

One should point out that fixing local gauge invariance as presented here does
not imply that the resulting expressions will automatically provide good
results for the problem at hand. It merely means that whatever is missing for a
good description will not be due to violation of local gauge invariance. In
other words, anything that should be found lacking as far as reproducing of
data is concerned would necessarily be resulting from manifestly transverse
current mechanisms not relevant for local gauge invariance.

The locally gauge-invariant Reggeization procedure outlined here is currently
being applied to describe Jefferson Lab data~\cite{clas14} for $\gamma + n \to
K^+ +\Sigma^*(1385)^-$ at photon energies between 1.5 and 2.5 GeV. The
preliminary results are encouraging; the full report will be published
elsewhere~\cite{WHH2015}.

Finally, we mention once more that the procedure given here can also be used
for the Reggeization of the $u$-channel in terms of baryonic Regge
trajectories. With the details given here, it should be quite obvious how to
implement this for the $u$-channel in a locally gauge-invariant manner (see
also footnote \ref{foot:schannel}).

%\acknowledgments
\section*{Acknowledgment}
One of the authors (H.H.) gratefully acknowledges discussions with Kanzo
Nakayama. J.H. acknowledges partial support by the Major State Basic Research
Development Program in China under grant 2014CB845405 and the National Natural
Science Foundation of China under grant 11275235.

\appendix

%%%%%%%%%%%%%%%%%%%%%%%%%%%%%%%%%%%%%%%%%%%
%%%%%%%%%%%%%%%%%%%%%%%%%%%%%%%%%%%%%%%%%%%
%%%%%%%%%%%%%%%%%%%%%%%%%%%%%%%%%%%%%%%%%%%
\section{Generic minimal interaction current}\label{sec:app}

To make the present paper self-contained, we provide in this Appendix the
generic expression for the \textit{minimal} interaction current necessary for
preserving local gauge invariance in the process (\ref{eq:process}). We largely
follow here Ref.~\cite{hh06}, but we provide additional clarification about
constraints on the parameter $\hat{h}$ introduced in Ref.~\cite{hh06}. Also, as
it is sufficient for the present purpose, we assume on-shell kinematics (for
the off-shell case, see Ref.~\cite{hh06}). The variables used in the following
are those of Fig.~\ref{fig:Msutc}.

The minimal interaction current appropriate for the hadronic vertex of the form
(\ref{eq:modelvertex}) reads~\cite{hh06}
\begin{equation}
  M_\IC^\mu = m_c^\mu\,f_t + \hF(q)\,C^\mu~,
  \label{eq:MintGeneric}
\end{equation}
where $m_c^\mu$ is a Kroll-Ruderman-type bare contact current resulting from an
elementary four-point Lagrangian appropriate for the reaction under
consideration. The effect of this current is to make the photoprocess of
Fig.~\ref{fig:Msutc} locally gauge invariant if all scalar form factors $f_x$
are put to unity. The auxiliary current $C^\mu$ is given by~\cite{hh06}
\begin{align}
  C^\mu &=-
  e_m(2q-k)^\mu\frac{f_t-1}{t-M_m^2}\big(\delta_s f_s + \delta_uf_u- \delta_s\delta_u f_sf_u\big)
  \nonumber\\[1ex]
  &\mbox{}\quad
  -e_{b'}(2p'-k)^\mu\frac{f_u-1}{u-M_{b'}^2}\big(\delta_t f_t + \delta_sf_s- \delta_t\delta_s f_tf_s\big)
  \nonumber\\[1ex]
  &\mbox{}\quad
  -e_{b}(2p+k)^\mu\frac{f_s-1}{s-M_b^2}\big(\delta_u f_u + \delta_tf_t- \delta_u\delta_t f_uf_t\big)
 \nonumber\\[1ex]
 &\mbox{}\quad+\hat{A}(s,u,t)(1-\delta_s f_s)(1-\delta_u f_u)(1-\delta_t f_t)
  \nonumber\\[1ex]
  &\mbox{}\quad\times
  \Bigg[
  e_m\frac{(2q-k)^\mu}{t-M_m^2}+e_{b'}\frac{(2p'-k)^\mu}{u-M_{b'}^2}
  +e_{b}\frac{(2p+k)^\mu}{s-M_b^2}  \Bigg],
 \label{eq:CauxGeneric}
\end{align}
where the factors $\delta_x$ ($x=s,u,t$) are unity if the corresponding channel
contributes to the reaction in question, and zero otherwise.
%  \footnote{The terms
%    with products $\delta_x\delta_y$ ($x,y=s,u,t$) only contribute if all three
%    hadron legs carry non-zero charge which can happen only if one of the baryons
%    is doubly charged (e.g., as in the process $\Delta^{++}\to p+\pi^+$); these terms can be omitted
%    otherwise.}
This contact current is manifestly nonsingular since the form factors become
unity at the respective poles thus providing well-defined $0/0$ situations. The
function $\hat{A}(s,u,t)$ is an arbitrary (complex) phenomenological function,
possibly subject to crossing symmetry constraints, that must vanish at high
energies. The expression here follows from Eq.~(31) of Ref.~\cite{hh06}
choosing the function $\hat{h}$ appearing there as $\hat{h}=1-\hat{A}$. The
vanishing high-energy limit of $\hat{A}$ is necessary to prevent the
``violation of scaling behavior" noted in Ref.~\cite{DrellLee1972} if $\hat{h}$
is different from unity at high energies.

The $\hat{A}$-dependent term in Eq.~(\ref{eq:CauxGeneric}) is easily seen to be
manifestly transverse in view of the charge-conservation relation
(\ref{eq:conscharge}) and therefore not necessary for preserving local gauge
invariance. However, it provides added flexibility when fitting data. In
principle, of course, any transverse (nonsingular) current may be added to the
right-hand side of (\ref{eq:MintGeneric}) without affecting gauge invariance.

The four-divergence of $C^\mu$ evaluates to
\begin{equation}
  k_\mu C^\mu = e_m\, f_t + e_{b'}\, f_u - e_b\, f_s~.
\end{equation}
In deriving this result repeated use was made of the charge-conservation
relation (\ref{eq:conscharge}). This is the scalar form of the generalized
Ward-Takahashi identity~(\ref{eq:gWTIint}) for the interaction current. The
right-hand side here vanishes for structureless particles where all form
factors are replaced by unity.

For the entire interaction current (\ref{eq:MintGeneric}) one then finds
\begin{align}
  k_\mu M^\mu_\IC &= \left[k_\mu m_c^\mu + e_m\,\hF(q)\right]f_t
  \nonumber\\
  &\mbox{}\qquad
  +e_{b'}\,\hF(q) f_u - e_b\,\hF(q) f_s~.
  \label{eq:kmuMmu}
\end{align}
Since the structureless contact current $m_c^\mu$ also must satisfy the gWTI
(\ref{eq:gWTIint}) with all form factors replaced by unity, we have
\begin{align}
k_\mu m_c^\mu &= e_m\hF(q-k)  +e_{b'}\hF(q)- e_{b}\hF(q)
   \nonumber\\
   &= e_m \hF(q-k)  -e_m\hF(q)~.
   \label{eq:gWTIintbare}
\end{align}
Equation~(\ref{eq:kmuMmu}) then reads
\begin{align}
  k_\mu M^\mu_\IC &=  e_m\,\hF(q-k) f_t
  \nonumber\\
  &\mbox{}\qquad
  +e_{b'}\,\hF(q) f_u - e_b\,\hF(q) f_s~,
\end{align}
which is the full gWTI  (\ref{eq:gWTIint}) with structure for the vertex of the
form (\ref{eq:modelvertex}).

\subsection{Beyond model treatment}\label{app:beyond}
We mention without going into much detail here that one may generalize the
vertex (\ref{eq:modelvertex}) by employing an expansion of the form
\begin{equation}
  F(p_{b'},p_b) = \iso\sum_i\lambda_i  \hF_i(q_m) f_i(q_m^2,p_{b'}^2,p_b^2)~,
  \label{eq:genvertex}
\end{equation}
where the sum extends over all possible coupling operators $\hF_i$, each with
its own (normalized) form factor $f_i$ such that the mixing parameters
$\lambda_i$ add up to unity,
\begin{equation}
  \sum_i\lambda_i =1~.
\end{equation}
In principle, one may even consider a formulation where the scalar functions
$f_i$ are no longer phenomenological suppression functions, but are determined
within a consistent dynamical framework. Depending on the sophistication of
this framework, the expansion (\ref{eq:genvertex}) then may be made arbitrarily
close to a complete dynamical description for the three-point vertex $b\to
m+b'$.

To accommodate the combination vertex (\ref{eq:genvertex}) with more than one
coupling operators $\hF_i$, one needs bare currents $m_{c,i}^\mu$ for each one
satisfying a separate gWTI like (\ref{eq:gWTIintbare}) and leading to, in
particular,
\begin{equation}
  k_\mu m_{c,i}^\mu = e_m\hF_i(q-k)-e_m \hF_i(q)~.
\end{equation}
The analog of the interaction current ansatz (\ref{eq:MintGeneric}) then is
\begin{equation}
  M^\mu_\IC = \sum_i \lambda_i\left[m_{c,i}^\mu f_{i,t}+\hF_i(q) C^\mu_i\right]~,
\end{equation}
where $f_{i,t}$ is the form factor $f_i$ in the $t$-channel and $C_i^\mu$ is
like (\ref{eq:CauxGeneric}) using $f_i$ alone. Everything then goes through as
before, and so this current obviously satisfies the correct gWTI for the vertex
(\ref{eq:genvertex}) and thus preserves local gauge invariance by construction.

%%%%%%%%%%%%%%%%%%%%%%%%%%%%%%%%%%%%
%%%%%%%%%%%%%%%%%%%%%%%%%%%%%%%%%%%%
%%%%%%%%%%%%%%%%%%%%%%%%%%%%%%%%%%%%

\end{document}